\documentstyle[12pt]{article}

\advance \textheight by \topskip
\begin{document}

\title{Gauge Dependence of Four-Fermion QED
Green Function and  Atom-Like Bound State
Calculations}
\author{Grigorii B. Pivovarov\thanks{e-mail address:
gbpivo@ms2.inr.ac.ru}\\
Institute for Nuclear Research \\
of the Russian Academy of Sciences,
Moscow 117312, Russia}
\maketitle
\begin{abstract}
We derive a reation between four-fermion QED
Green functions of different covariant gauges
which defines the gauge dependence completely.
We use the derived gauge dependence to check the
gauge invariance of atom-like bound state
calculations. We find that the existing QED procedure
does not provide gauge invariant binding energies.
A way to a corrected gauge invariant procedure is pointed out.
\end{abstract}

\section{Introduction}

The persisting discrepancy between theory and experiment for
positronium width \cite{Westbrook} is a chalenge for QED. At the moment
the hope is on taking into account corrections of relative order
$\alpha^{2}$ \cite{Lepage,Khriplovich}. In the circumstances the
question of self-consistency of the calculations, in particular,
of gauge invariance of the result is of prime concern.

The modern way to calculate parameters of two-particle
atom-like bound states is to extract them from corresponding
four-fermion QED Green function (see, for example, \cite{Lepage78,%
Remiddi,Steinman} and this paper below). Thus, to check the gauge
independence of the calculated bound state parameters, one should carry
the gauge parameter through all the extraction procedure.
(An example of this see in \cite{Adkins} where the gauge independence
of the correction to the positronium width of relative
order $\alpha$ was checked.) The extraction procedure
gets more and more complicated with an increase in order
of radiative corrections and  direct order by order check of
gauge invariance becomes impractical as a check of self-consistency
of the calculations. Instead, one would like to exploit gauge invariance
choosing a most convenient gauge and switching from one gauge to another
in the process of the calculations. In view of these complications,
it seems pertinent to make a step out of the concrete practice of
bound-state calculations and to study first the gauge dependence of the
four-fermion QED Green function itself without taking into account
the complications of the bound-state parameter calculations.

In the present paper we derive a relation between four-fermion QED
Green functions of different values of gauge-fixing parameter
(we consider the covariant gauges only). The relation completely
defines the evolution of the Green function in the gauge-fixing
parameter. Our derivation does not use perturbation theory.
Next, we use our relation to check gauge invariance of the extraction
procedure of atom-like bound-state parameters. The result is
negative. It turns out that the existing procedure provides
gauge-dependent answers for binding energies. We find a
flaw in the procedure which is responsible for the gauge-dependence
of the result and point the way to its correction.

Next section contains a derivation of the evolution in the gauge-fixing
parameter; section 3 comprises a brief recall of the extraction procedure
and an utilization of the general evolution formula from section 2
for an analysis of gauge-dependence of the extraction;
in the last, fourth, section we point out the reason for the gauge
dependence and the way to the correct procedure.

\section{Evolution in Gauge-Fixing Parameter}

Let us consider the four-fermion QED Green function
\begin{equation}
\label{Gf}
G_{\beta}(x_{f},\overline{x}_{f},x_{i},\overline{x}_{i})\equiv
        i\int D\psi DA\, \exp\left(iS_{QED}(\beta)\right)
        (\overline{\psi}(\overline{x}_{f}) \psi(x_{f}))
        (\overline{\psi}({x}_{i}) \psi(\overline{x}_{i}))\, ,
\end{equation}
where $x_{f}(\overline{x}_{f})$ is a coordinate of outgoing
particle (antiparticle) and $x_{i}(\overline{x}_{i})$ is the
same for ingoing pair. The definition of
gauge fixing parameter $\beta$ is given by corresponding photon
propagator:
\begin{equation}
\label{gfix}
D_{\mu \nu}(\beta,x) = \int\frac{dk}{(2\pi)^{4}}
    \left(-g_{\mu \nu} + \beta\frac{k_{\mu}k_{\nu}}{k^{2}}\right)
        \frac{i}{k^{2}}e^{ikx}.
\end{equation}

Our aim is to study the dependence of $G_{\beta}$ on $\beta$.
To this end, it is useful to consider a Green function
in external photon field, $G(A)$, which is a
result of integration over the fermion field in the rhs of (\ref{Gf}).
{}From the one hand, it is simply connected to the Green function
\cite{Vass}:
\begin{equation}
\label{connection}
G_{\beta} = (e^{L_{\beta}}G(A))_{A=0}\,,\;
L_{\beta}\equiv\frac{1}{2}\frac{\delta}{\delta A_{\mu}}D_{\mu \nu}(\beta)
                       \frac{\delta}{\delta A_{\nu}}.
\end{equation}
(In this formula each $L_{\beta}$ generates a photon propagator;
the dependence
on the coordinates of ingoing and outgoing particles is suppressed
for brevity.)
{}From the other hand, $G(A)$ is siply connected to a gauge invariant object
$G_{inv}(A)$:
\begin{equation}
\label{coninv}
G(A) = G_{inv}(A) \exp\left(ie\int^{x_{f}}_{\overline{x}_{f}}A_{\mu}dx^{\mu}
                       -ie\int^{x_{i}}_{\overline{x}_{i}}A_{\mu}dx^{\mu}
                       \right).
\end{equation}
The gauge invariance of $G_{inv}$ means that it is independent of
the longitudinal component of $A$:
\begin{equation}
\label{gi}
\partial_{\mu}\frac{\delta}{\delta A_{\mu}}G_{inv}(A) = 0
\end{equation}
and is a consequence of gauge invariance of the combination
\begin{equation}
\overline\psi(x)\exp\left(ie\int^{x}_{y}A_{\mu}dz^{\mu}
                     \right)\psi(y).
\end{equation}

A substitution of (\ref{coninv}) into (\ref{connection}) yields
\begin{equation}
\label{hot}
G_{\beta} = \left (e^{L_{\beta}}G_{inv}(A)
\exp\left(ie\int^{x_{f}}_{\overline{x}_{f}}A_{\mu}dx^{\mu}
     -ie\int^{x_{i}}_{\overline{x}_{i}}A_{\mu}dx^{\mu}\right)\right)_{A=0}.
\end{equation}
Let us take a $\beta$-derivative of both sides of this equation:
\begin{equation}
\label{almost eq}
\frac{\partial}{\partial\beta}G_{\beta} =
            \left (e^{L_{\beta}}(\partial_{\beta}L_{\beta})G_{inv}(A)
\exp\left(ie\int^{x_{f}}_{\overline{x}_{f}}A_{\mu}dx^{\mu}
     -ie\int^{x_{i}}_{\overline{x}_{i}}A_{\mu}dx^{\mu}\right)\right )_{A=0} .
\end{equation}
To get an evolution equation, one needs to express the rhs of
this equation in terms of $G_{\beta}$. It is possible because
$(\partial_{\beta}L_{\beta})$ commutes  with $G_{inv}(A)$ and gives a
$c$-factor when acts on the consequent exponential. So, (\ref{almost eq})
transforms itself into
\begin{equation}
\label{equation}
\frac{\partial}{\partial\beta}
G_{\beta}(x_{f},\overline{x}_{f},x_{i},\overline{x}_{i}) =
        F(x_{f},\overline{x}_{f},x_{i},\overline{x}_{i})
        G_{\beta}(x_{f},\overline{x}_{f},x_{i},\overline{x}_{i}),
\end{equation}
where we have restored the $x$-dependence
and used $F$ to denote the action of
$(\partial_{\beta}L_{\beta})$ on the exponential:
\begin{eqnarray}
\label{F-def}
\lefteqn{(\partial_{\beta}L_{\beta})
\exp\left(ie\int^{x_{f}}_{\overline{x}_{f}}A_{\mu}dx^{\mu}
     -ie\int^{x_{i}}_{\overline{x}_{i}}A_{\mu}dx^{\mu}
     \right) \equiv}\nonumber \\
& &       F(x_{f},\overline{x}_{f},x_{i},\overline{x}_{i})
\exp\left(ie\int^{x_{f}}_{\overline{x}_{f}}A_{\mu}dx^{\mu}
     -ie\int^{x_{i}}_{\overline{x}_{i}}A_{\mu}dx^{\mu}
     \right).
\end{eqnarray}

An explanation is in order: In deriving (\ref{equation}) we
have used a commutativity of $(\partial_{\beta}L_{\beta})$
and $G_{inv}(A)$; it is a direct consequence of gauge invariance
of $G_{inv}$ (see (\ref{gi})) and the fact that
$(\partial_{\beta}L_{\beta})$ contains only derivatives in
longitudinal components of $A$ (see (\ref{connection}) for a
definition of $L_{\beta}$ and (\ref{gfix}) for $\beta$-dependence
of $D_{\mu \nu}$).

The solution of eq.(\ref{equation}) for $\beta$-evolution is
\begin{equation}
\label{solution}
G_{\beta}(x_{f},\overline{x}_{f},x_{i},\overline{x}_{i}) =
    \exp\left((\beta-\beta_{0})
    F(x_{f},\overline{x}_{f},x_{i},\overline{x}_{i})
        \right)
        G_{\beta_{0}}(x_{f},\overline{x}_{f},x_{i},\overline{x}_{i}).
\end{equation}

To get the final answer one needs an explicite view of $F$
from (\ref{solution}). It is easily deduced from the $F$-definition
(\ref{F-def}) and the following representation for the longitudinal
part of the photon propagator:
\begin{equation}
\label{representation}
\partial_{\beta}D_{\mu \nu}(\beta,x) =
     -\frac{1}{16\pi^{2}}\partial_{\mu}\partial_{\nu}
     \ln((x^{2}-i\varepsilon)m^{2}),
\end{equation}
where $m$ is an arbitrary mass scale which is fixed, for
defineteness, on the fermion mass. Then, up to an additive
constant,
\begin{equation}
\label{repres}
F = \frac{\alpha}{4\pi}\left(
 \ln\frac{1}{m^{4}(x_{f}-\overline{x}_{f})^{2}(x_{i}-\overline{x}_{i})^{2}}
+\ln\frac{(x_{f}-x_{i})^{2}(\overline{x}_{f}-\overline{x}_{i})^{2}}
         {(x_{f}-\overline{x}_{i})^{2}(\overline{x}_{f}-x_{i})^{2}}
                      \right).
\end{equation}

Substituting (\ref{repres}) into (\ref{solution}), we get
our final aswer for $\beta$-evolution:
\begin{eqnarray}
\label{answer}
G_{\beta}(x_{f},\overline{x}_{f},x_{i},\overline{x}_{i})&=&
 \left[
      \frac{Z(x_{f}-x_{i})^{2}(\overline{x}_{f}-\overline{x}_{i})^{2}}
           {m^{4}(x_{f}-\overline{x}_{f})^{2}(x_{i}-\overline{x}_{i})^{2}
          (x_{f}-\overline{x}_{i})^{2}(\overline{x}_{f}-x_{i})^{2}}
    \right]^{\frac{\alpha}{4\pi}(\beta-\beta_{0})} \times\nonumber\\
&&G_{\beta_{0}}(x_{f},\overline{x}_{f},x_{i},\overline{x}_{i}) .
\end{eqnarray}
The normalization $Z$ is infinite before the ultraviolet
renormalization. After the renormalization it is scheme-dependent
and calculable order by order in perturbation theory. We
will not need its value in what follows.

\section{The Bound State Parameters And The Four-Fermion QED
Green Function}

The four-fermion QED Green function contains too much
information for one who just going to calculate bound-sate parameters.
Ona can throw away unnessesary information by putting
senter of mass space-time coordinate of ingoing pair
and relative times of both ingoing and outgoing pairs to zero:
\begin{equation}
\label{eqtimes}
G_{(et) \beta}(t,{\bf x},{\bf r'},{\bf r})\equiv
G_{\beta}\left(x_{f}(t,{\bf x},{\bf r'}),
               \overline{x}_{f}(t,{\bf x},{\bf r'}),
               x_{i}({\bf r'}),
               \overline{x}_{i}({\bf r'})
          \right),
\end{equation}
where the space-time coordinates depend on a space-time
coordinate of the center of mass of the outgoing pair $(t,{\bf x})$ and
a relative space coordinate of outgoing $(\bf r')$ and ingoing $(\bf r)$
pair. In the case of equal masses
\begin{eqnarray}
\label{def r}
x_{f}=(t,{\bf x}+\frac{{\bf r'}}{2}),&\;&
\overline{x}_{f}=(t,{\bf x}-\frac{{\bf r'}}{2}),\nonumber \\
x_{i}=(0,\frac{{\bf r}}{2}),&\;&
\overline{x}_{i}=(0,-\frac{{\bf r}}{2}).
\end{eqnarray}

$G_{(et)\beta}$ still contains an unnecessary piece of
information --- the dependence on the center of mass
space coordinate. The natural way to remove it is
to go over to momentum representation and put the
center of mass momentum to zero. In coordinate representation,
which is more convenient for gauge invariance check, we define
the propagator $D_{\beta}$ of the fermion pair:
\begin{equation}
\label{propDef}
G_{(et)\beta}(t,{\bf x},{\bf r'},{\bf r}) \equiv
D_{\beta}(t,{\bf r'},{\bf r})\delta({\bf x}) + \ldots,
\end{equation}
where dots denote terms with derivatives of $\delta({\bf x})$.
It is natural to consider $D_{\beta}$ as a time dependent kernel
of an operator acting on wave-functions of relative coordinate.
In what follows we will not make difference between a kernel
and the corresponding operator. The naturalness of the
above definition of the propagator is apparent in the
nonrelativistic approximation:
\begin{equation}
\label{NR}
{e^{i2mt}}D_{\beta}(t) \approx
                \sum_{E_{0}} \theta(t)e^{-iE_{0}t} P(E_{0}),
\end{equation}
where the summation runs over the spectrum of nonrelativistic
Coulomb problem and $P(E_{0})$ are the projectors onto corresponding
subspaces of the nonrelativistic state space. One can obtain
(\ref{NR}) keeping leading term in $\alpha$-expansion of
the lhs if one will keep $t\propto 1/\alpha^{2}$ and
${\bf r'},{\bf r}\propto 1/\alpha$ (see \cite{Steinman,Pivovarov}).
The subscript on $E_{0}$ is to denote that it will get
radiative corrections (see below). The exponential in the lhs
is to make a natural shift in energy zero.
In what follows we will
include the energy shift in the definition of $
D_{\beta}(t)$.

The next step in calculation of radiative corrections to the energy
levels is a crucial one: one should make an assumption about
the general form of a deformation of the $t$-dependence of
the rhs of (\ref{NR}) caused by relativistic corrections.
A naturall guess and the one which leads to the generally accepted
rules of calculation of the relativistic corrections to the energy
eigenvalues (see, for example \cite{Lepage78}) is to suppose that one can
contrive oscillating part of the exact propagator $D_{\beta}$
from the rhs of (\ref{NR}) just shifting energy levels and
modifying the operator coefficiens $P(E_{0})$:
\begin{equation}
\label{guess}
D_{\beta}(t) = \sum_{E_{0}+\Delta_{E_{0}}} \theta(t)
                        e^{-i\left(E_{0}+\Delta_{E_{0}}\right)t}
                        P_{\beta}(E_{0}+\Delta_{E_{0}}) + \ldots,
\end{equation}
where dots denote terms which are slowly-varying in time
(the natural time-scale here is $1/E_{0}$). The additional
subscript $\beta$ on $P_{\beta}$ is to denote that
oscillating part of $D_{\beta}(t)$ can acquire a gauge parameter
dependence from relativistic corrections.

Let us see how one can use eq.(\ref{guess}) in energy level
calculations. It is quite sufficient to consider $D_{\beta}(t)$
on relatively short times when $\Delta_{E_{0}} t\ll 1,\, E_{0}t\sim 1$.
For such times one can approximate $D_{\beta}$
expanding the rhs of eq.(\ref{guess}) over $\Delta_{E_{0}}t$:
\begin{equation}
\label{simple}
D_{\beta}(t) \approx \sum_{E_{0}} \theta(t)e^{-iE_{0}t}
                  \sum_{k}t^{k}A^{(k)}_{\beta}(E_{0}),
\end{equation}
where
\begin{equation}
\label{AE}
A^{(k)}_{\beta}(E_{0}) = \sum_{\Delta_{E_{0}}}
         \frac{(-i\Delta_{E_{0}})^{k}}{k!}P_{\beta}(E_{0}+\Delta_{E_{0}}).
\end{equation}
An extraction of these objects from the perturbation
theory is an interim step in the level shift calculations.
(Here we should mention that in calculational practice
$A^{(k)}_{\beta}(E_{0})$ are
exracted in momentum representation --- i.e.  not as coefficients
near the powers of time but as the ones near the propagator-like
singularities $(E-E_{0}+i\varepsilon)^{-(k+1)}$.)
To come nerer to the level shift values, useful objects are
\begin{equation}
\label{A}
A^{(k)}_{\beta} \equiv \sum_{E_{0}}A^{(k)}_{\beta}(E_{0})i^{k}k!.
\end{equation}
Namely, as notations of (\ref{AE}) suggest, eigenvalues
of $A^{(0)}_{\beta}$ should be equal to normalizations of bound state
wave functions which are driven from unit by relativistic
corrections while the eigenvalues of $A^{(k)}_{\beta}$ should
be energy shifts to the $k$-th power times corresponding normalizations.
Thus, the eigenvalues of
\begin{equation}
\label{Skdef}
S^{(k)}_{\beta} \equiv \frac{\left[A^{(0)}_{\beta}\right]^{-1}A^{(k)}_{\beta}
                + A^{(k)}_{\beta}\left[A^{(0)}_{\beta}\right]^{-1}}{2}
\end{equation}
should be just energy shifts to the $k$-th power. Thus, we define
\begin{equation}
\label{Sdef}
S_{\beta} \equiv S_{\beta}^{(0)}
\end{equation}
to be the energy shift operator: its eigenvalues are the energy
level shifts caused by relativistic corrections.
Our aim is now to check $\beta$-independence of $S_{\beta}$
eigenvalues.

Some notes are in order: If the conjecture (\ref{guess})
is true, $A^{(0)}_{\beta}$ should commute with $S^{(k)}_{\beta}$
and the following relation should hold:
\begin{equation}
\label{powerrel}
S^{(k)}_{\beta} = \left[S_{\beta}\right]^{k}
\end{equation}
This relation was suggested as a check of the
cojecture (\ref{guess}) in \cite{Steinman} and, to
our knowlege, has never been checked. Another thing to note is
that relativistic corrections affects the form
of the scalar product of wave functions and, thus, one
shoud add a definition of operator products to
the formal expressions (\ref{Skdef}),(\ref{powerrel}).
But the level of accuracy to which we will operate
permits us not to go into this complication and use
the operator products as they are in the nonrelativistic
approximation --- i.e. as the convolution
of the corresponding kernels.

The way to the gauge invariance check of the energy shift calculations
is clear now: Using the gauge evolution relation (\ref{answer})
one should find the $\beta$-dependence of $S_{\beta}$ and then
of its eigenvalues. As $S_{\beta}$ is defined
in (\ref{Sdef}),(\ref{Skdef}) through $A^{(k)}_{\beta}$'s which are,
in turn, defined in (\ref{simple})
through the propagator $D_{\beta}$, the first step is to simplify
(\ref{answer}) to the reduced case of zero relative time and
total momentum of the fermion pair:
\begin{eqnarray}
\label{reduced}
D_{\beta}(t,{\bf r'},{\bf r})&=&\left[
                        \frac{\left(1-({\bf r'}-{\bf r})^{2}/(4t^{2})
                             \right)}
                            {\left(1-(({\bf r'}+{\bf r})^{2}/(4t^{2})
                             \right)}
                              \right]^
               {\frac{\alpha}{2\pi}(\beta-\beta_{0})}\times\nonumber \\
                             & &\left[
                        \frac{Z}
                            {m^{2}{\bf r'}^{2}m^{2}{\bf r}^{2}}
                              \right]^{\frac{\alpha}{4\pi}(\beta-\beta_{0})}
                             D_{\beta_{0}}(t,{\bf r'},{\bf r}).
\end{eqnarray}
The factor in the square brackets of the second line is time-independent
and futher factorizible on factors depending on either ingoing or outgoing
pair parameters. This reduce the influence of this factor to a change
in the normalization of states. Being interested in
gauge invariance of energy shifts, we omit this factor in what follows.
Let us turn to the analysis of the influence of the factor in the first
line of (\ref{reduced}).

This factor is close to unit in the atomic scale
${\bf r'},{\bf r}\sim 1/\alpha,\,t\sim1/\alpha^{2}$.
We will use its approximate form:
\begin{equation}
\label{approx}
Factor \approx 1 + \frac{\alpha}{2\pi}(\beta-\beta_{0})
                \frac{{\bf r'}{\bf r}}{t^2} + O(\alpha^{5}).
\end{equation}

One can read the dependence of $A^{(k)}_{\beta}$
on $\beta$ from (\ref{simple}),(\ref{reduced}),(\ref{approx})
as
\begin{equation}
\label{betadep}
A^{(k)}_{\beta} \approx A^{(k)}_{\beta_{0}} -
               \frac{\alpha}{2\pi}\frac{(\beta-\beta_{0})}{(k+1)(k+2)}
               {\bf r}A^{(k+2)}_{\beta_{0}}{\bf r},
\end{equation}
where $\bf r$ is the vector operator of relative position
of interacting particles.
The mixing of different $A^{(k)}_{\beta}$'s with a change in
the gauge parameter is due to the presence of $1/t^{2}$ in
the rhs of (\ref{approx}).
Finally, using the definition
(\ref{Sdef}), relations (\ref{powerrel}) and the fact that
\begin{equation}
\label{unit}
A^{(0)} \approx 1
\end{equation}
in the nonrelativistic approximation one can derive the
following $\beta$-dependence of $S_{\beta}$:
\begin{eqnarray}
\label{Sanswer}
S_{\beta}&\approx&S_{\beta_{0}} -\nonumber \\
         &       &\frac{\alpha}{2\pi}(\beta-\beta_{0})
                  \left(\frac{1}{6}{\bf r}S_{\beta_{0}}^{3}{\bf r} -
                   \frac{1}{4}S_{\beta_{0}}{\bf r}S_{\beta_{0}}^{2}{\bf r} -
                   \frac{1}{4}{\bf r}S_{\beta_{0}}^{2}{\bf r}S_{\beta_{0}}
                   \right).
\end{eqnarray}
Treating the term in the last line of the rhs of the above relation
as a perturbation, one can get an approximate
value of the $\beta$-dependent
piece of the energy shift just averaging the perturbation
with respect to the corresponding eigenstate of $S_{\beta_{0}}$.

Thus, we get for the leading order of $\beta $-derivative of
an energy shift the following representation:
\begin{equation}
\label{leading}
\left(\frac{\partial}{\partial\beta}\Delta_{\beta}\right)_{L}=
               -\frac{\alpha}{2\pi}
                  \left(\frac{1}{6}\left\langle
                  {\bf r}S_{L}^{3}{\bf r}\right\rangle -
  \frac{1}{4}\left\langle S_{L}{\bf r}S_{L}^{2}{\bf r}\right\rangle -
  \frac{1}{4}\left\langle{\bf r}S_{L}^{2}{\bf r}S_{L}\right\rangle
                   \right),
\end{equation}
where $\langle\ldots\rangle$ means averaging with respect to
the corresponding nonrelativistic eigenstate and the subscript
$L$ means the leading  order in $\alpha$-expansion.

Eq.(\ref{leading}) is sufficient to define an order in
$\alpha$ in which the energy shifts become gauge dependent:
\begin{equation}
\label{order}
\left(\frac{\partial}{\partial\beta}\Delta_{\beta}\right)_{L}
                        \sim \alpha^{11}.
\end{equation}
Here we have taken into account that ${\bf r}\sim1/\alpha$
and $S_{L}\sim\alpha^{4}$.

To have a gauge dependence in any observable is clearly
unacceptable. In the next section we will see how one
should correct the above procedure of energy shift
extraction from the QED Green function to get rid of
the gauge dependence of energy shifts.

\section{A Way Out}

The procedure recalled in the previous section
is based on the conjecture (\ref{guess}).
A consequence of this conjecture is the
gauge dependence of energy shifts of (\ref{leading}).
One can conclude that the conjecture is wrong.
In particular, as one can infer from eq.(\ref{reduced}),
the operator coefficients near the oscillating exponentials
in (\ref{guess}) shoud get a time dependence from relativistic
corrections. Even if in some gauge they are time independent,
the gauge parameter evolution should generate a dependence
which in the leading order in $\alpha$ reduce itself to
the following replacement in (\ref{guess}):
\begin{equation}
\label{replacement}
P_{\beta}(E_{0}+\Delta_{E_{0}})\rightarrow
P_{\beta}(E_{0}+\Delta_{E_{0}}) + \frac{\Sigma_{\beta}(E_{0})}{t^{2}}.
\end{equation}
That $\Sigma_{\beta}(E_{0})$ has nothing to do with
energy shifts but will give contributions
to $A^{(k)}_{\beta}(E_{0})$'s from eq.(\ref{simple}).
Being gauge dependent these contributions lead to the
gauge dependence of energy shifts.

The way to the correct procedure is to through away
terms like $\Sigma_{\beta}(E_{0})/t^{2}$ prior to the definition of
the energy shift operator. Thus, a necessary step
in the process of extracting energy shifts from
the QED Green function (and the one
which necessity is not recognized in the stanard procedure)
is to calculate and subtract contributions like the last
term in the rhs of (\ref{replacement}) from the propagator
of the fermion pair.

Below we report on a calculation of $\Sigma_{\beta}(E_{0})$ from
(\ref{replacement}). The most economical way to calculate it
is to note that the energy dependence
of the Fourier transform of the corresponding
contribution to the propagator is
\begin{equation}
\label{fourier}
(E-E_{0})\ln(-(E-E_{0}+i\varepsilon))
\end{equation}
and that it comes from diagrams describing radiation
and subsequent absorption of a soft photon with no change
in the level $E_{0}$ of the radiating and absorbing bound state.
Similar contributions (with another power of energy before the $log$)
are well known for the propagator of a charged fermion \cite{Lifshits}

The first step in our calculation is to
present the pair propagator in the following form:
\begin{equation}
\label{soft}
D_{\beta}(t)\approx\left(e^{L_{s}}e^{ie{\bf rA}(t)}D_{inv}(t,A)
                       e^{-ie{\bf rA}(0)}
               \right)_{A=0},
\end{equation}
where $L_{s}$ is the same as in (\ref{connection}) except a
restriction on the momentum of photon propagator ---
the range of its variation is restricted to the soft
region which border is of order of atomic
binding energies; the exponentials with
gauge potential are originated from
the ones in (\ref{hot}); $D_{inv}$ is a descendant of $G_{inv}$
from (\ref{hot}): to go over from $G_{inv}$ to $D_{inv}$
one should make all pairing of non-soft photons in $G_{inv}$
and all the reductions of space-time coordinats which was involved
in going over from the $G_{\beta}$ of (\ref{Gf}) to the $D_{\beta}$
of (\ref{propDef}); at last, all gauge potentials in (\ref{soft})
are taken at zero of space coordinate in accord with the
$\delta({\bf x})$ of eq.(\ref{propDef}). The difference
between the lhs and the rhs  of eq.(\ref{soft}) does not conribute
to the term under the calculation.

The leading in the nonrelativistic approximation contribution to
$D_{inv}$ is the same as for $D_{\beta}$ --- it is just
the propagator of the nonrelativistic Coulomb problem. We explicitly
calculate the leading contribution
to the dependence of $D_{inv}(t,A)$ on the gauge potential
in its expansion over soft momenta of the external photons.
Not surprisingly, the dipole interaction of the pair with
the external photon field arise in this approximation:
\begin{equation}
\label{Adef}
D_{inv}(t,A) \approx \left(i\frac{\partial}{\partial t} - H_{c}
                                        + e{\bf r}{\cal E}(t)
                  \right)^{-1},
\end{equation}
where $H_{c}$ is the hamiltonian of the nonrelativistic
Coulomb problem and $\cal E$ is the strength of the electric field:
\begin{equation}
\label{Edef}
{\cal E}(t)\equiv -\dot{{\bf A}}(t) + \nabla A_{0}(t).
\end{equation}

Substituting (\ref{Adef}) in (\ref{soft}) and keeping terms
with only one soft photon propagator we get
expressions which sum contains the term under calculation:
\begin{equation}
\label{r1}
e^{2}\left(L_{s}
                {\bf rA}(t)D_{nr}(t){\bf rA}(0)\right)_{A=0},
\end{equation}
\begin{equation}
\label{r2}
e^{2}\left(L_{s}
                \int d\tau_{1}d\tau_{2}\,
                D_{nr}(t-\tau_{1}){\bf r}{\cal E}(\tau_{1})
                D_{nr}(\tau_{1}-\tau_{2}){\bf r}{\cal E}(\tau_{2})
                D_{nr}(\tau_{2})\right)_{A=0} ,
\end{equation}
\begin{eqnarray}
\label{r3}
ie^{2}\biggl(L_{s}
                \int d\tau\,\bigl(
          D_{nr}(t-\tau){\bf r}{\cal E}(\tau)D_{nr}(\tau){\bf rA}(0)&-&
          \\
&  &      {\bf rA}(t)D_{nr}(t-\tau){\bf r}{\cal E}(\tau)D_{nr}(\tau)
          \bigr)
                       \biggr)_{A=0},\nonumber
\end{eqnarray}
where $D_{nr}(t)$ is the propagator of the nonrelativistic Coulomb
problem from the rhs of eq.(\ref{NR}).

The next step is to pick out a contribution of a level $E_{0}$
in (\ref{r1}),(\ref{r2}),(\ref{r3}). That is achievable by the
replacement
\begin{equation}
\label{repl}
D_{nr}(t)\rightarrow e^{-iE_{0}t}\theta(t)P(E_{0}).
\end{equation}

The last ingredient that one needs to calculate
(\ref{r1}),(\ref{r2}),(\ref{r3}) is the time dependence of the soft
photon propagators. It can be deduced from (\ref{gfix})
as
\begin{eqnarray}
\label{time}
\left(L_{s}A_{i}(t_{1})A_{j}(t_{2})\right)&=&
                \theta\left((t_{1}-t_{2})^{2}>t_{c}^{2}\right)
 \frac{\delta_{ij}\left(-1+\frac{\beta}{2}\right)}{4\pi^{2}(t_{1}-t_{2})^{2}},
 \nonumber \\
\left(L_{s}A_{i}(t_{1}){\cal E}_{j}(t_{2})\right)&=&
                \theta\left((t_{1}-t_{2})^{2}>t_{c}^{2}\right)
 \frac{\delta_{ij}}{2\pi^{2}(t_{1}-t_{2})^{3}},\nonumber \\
\left(L_{s}{\cal E}_{i}(t_{1}){\cal E}_{j}(t_{2})\right)&=&
                \theta\left((t_{1}-t_{2})^{2}>t_{c}^{2}\right)
 \frac{\delta_{ij}}{\pi^{2}(t_{1}-t_{2})^{4}}.
\end{eqnarray}
Here the $\theta$-functions are to account for the softness of
the participating photons ($t_{c}\sim 1/E_{0}$).

Taking (\ref{time}) into account we get the
following contributions from
(\ref{r1}),(\ref{r2}),(\ref{r3}):
\begin{eqnarray}
\label{contr}
(\ref{r1})&\rightarrow& \frac{1}{t^{2}}\theta(t)e^{-iE_{0}t}
        \frac{\alpha}{\pi}\left(-1+\frac{\beta}{2}\right)
        {\bf r}P(E_{0}){\bf r},\nonumber \\
(\ref{r2})&\rightarrow& \frac{1}{t^{2}}\theta(t)e^{-iE_{0}t}
  \frac{\alpha}{\pi}\frac{2}{3}P(E_{0}){\bf r}P(E_{0}){\bf r}P(E_{0}),
  \nonumber \\
(\ref{r3})&\rightarrow& \frac{1}{t^{2}}\theta(t)e^{-iE_{0}t}
  \frac{\alpha}{\pi}i\left(P(E_{0}){\bf r}P(E_{0}){\bf r} -
  {\bf r}P(E_{0}){\bf r}P(E_{0})\right).
\end{eqnarray}

The sum of the above terms yields the result of our calculation:
\begin{eqnarray}
\label{sigmansw}
\Sigma_{\beta}(E_{0})&=&\frac{\alpha}{\pi}
\biggl( \frac{2}{3}P(E_{0}){\bf r}P(E_{0}){\bf r}P(E_{0}) +
(-1+\frac{\beta}{2}){\bf r}P(E_{0}){\bf r} +\nonumber\\
    & &  i(P(E_{0}){\bf r}P(E_{0}){\bf r} - {\bf r}P(E_{0}){\bf r}P(E_{0}))
\biggr) .
\end{eqnarray}

One can explicitly check that $\beta$-dependence of $\Sigma_{\beta}(E_{0})$
is the right one --- i.e. if one subtracts the $\Sigma$-term
from the propagator before the definition of the energy shift
operator, the latter becomes gauge independent. Another observation
is that the $\Sigma$-term cannot be killed by any choice
of the gauge (in contrast to the case of charged fermion propagator
where an analogous term is equal to zero in the Yennie gauge).

Summing up, in this paper we derived a relation between QED
Green functions of different gauges. We used it to check
the gauge invariance of the energy shift operator. It turns out
to be gauge dependent. This fact forced us to recognize
that energy shifts are not one, and the only one, source for
the positive powers of time near the oscillating exponentials
in the propagator of the pair. We found a particular
additional source of the positive powers of time which is
responsible for the gauge dependence of the naive
energy shift operator. We conclude by an observation
that at the moment we have not a  clear
definition of the energy shift operator --- to get
it one needs a criterion for picking out contributions
to the positive powers of time originating from the energy
shifts.

The author is grateful to A.~Kataev, E.~Kuraev, V.~Kuzmin, A.~Kuznetsov,
S.~Larin, Kh.~Nirov,
V.~Rubakov, D.~Son, P.~Tinyakov
for helpful discussions. This work was supported in
part by The Fund for Fundamental Research of Russia
under grant 94-02-14428.

\end{document}